\documentclass[twocolumn,showpacs,amsmath,amssymb,aps,prl,floats]{revtex4-1}
\usepackage{epsfig}
\usepackage{bm}
\usepackage{graphicx}
\usepackage{color}
\usepackage{hyperref}
\usepackage{float}
\usepackage{epstopdf}
\usepackage{wrapfig}

\begin{document}

\title{Understanding the multiwavelength observation of Geminga's TeV halo: the role of anisotropic diffusion of particles}
\author{Ruo-Yu Liu$^{1}$}\email{ruoyu.liu@desy.de}
\author{Huirong Yan$^{1,2}$}\email{huirong.yan@desy.de}
\author{Heshou Zhang$^{1,2}$}
\affiliation{$^1$Deutsches Elektronen Synchrotron (DESY), Platanenallee 6, D-15738 Zeuthen, Germany\\
$^2$School of Astronomy and Space Science, Nanjing University, Nanjing 210093, China\\
$^3$Institut f\"ur Physik und Astronomie, Universit\"at Potsdam, D-14476 Potsdam, Germany}

\begin{abstract}
In this letter we {\bf propose} that the X-ray and the TeV observations in the vicinity of Geminga can be understood in the framework of anisotropic diffusion of injected electrons/positrons. This interpretation only requires the turbulence in the vicinity of Geminga to be sub-Alfv{\'e}nic with the local mean magnetic field direction approximately aligned with our line of sight towards Geminga, without invoking extreme conditions for the environment, such as an extremely small diffusion coefficient and a weak magnetic field of strength $<1\mu$G as suggested in previous literature. 
\end{abstract}

\maketitle

{\it Introduction-} Recent observation of the High-Altitude Water Cherenkov Observatory (HAWC) has revealed a TeV gamma-ray halo around the Geminga pulsar, with a spatial extension of $\gtrsim 30$\,pc \citep{HAWC17_Geminga}. The TeV emission is believed to arise from cosmic-ray electrons/positrons (hereafter we do not distinguish positrons from electrons unless specified) injected from the pulsar wind nebula (PWN), via inverse-Compton (IC) scattering off cosmic microwave background (CMB) photons. The detection of such a diffuse TeV emission has been interpreted as the presence of a slow diffusion zone around the pulsar \citep{HAWC17_Geminga, Fang18, Profumo18, Xi18} in the framework of 1D isotropic diffusion. Under the same framework, Ref.\citep{Liu19} studied the X-ray observation by XMM-Newton and Chandra in the vicinity of Geminga and an upper limit of $5\times 10^{-15}\rm erg~cm^{-2}s^{-1}$ in $0.7-1.3\,$keV has been obtained for a region within $600''$ around the pulsar. {This translates to an upper limit for the magnetic field strength in the TeV halo, i.e., $\leq 0.8\mu$G, which is significantly weaker than the typical interstellar medium (ISM) magnetic field.} Furthermore, the combination of a small diffusion coefficient and a weak magnetic field would imply the saturation of the turbulence ($\delta B_g/B\simeq 1$ where $\delta B_g$ is the fluctuation amplitude of magnetic field at the gyro-scale of particles and $B$ is the mean magnetic field) and the Bohm limit of diffusion, which is, however, very difficult to achieve. For interstellar turbulence, the energy injection scale  is $\sim 100$pc and much larger than the gyroscale of the TeV-emitting electrons, where the resonant scattering happens. It is unlikely that $\delta B_g/B$ approaches to unity. { A plausible scenario} is the small-scale waves generated by instabilities. The electron flux at $~100$TeV is, nonetheless, too small to generate strong enough streaming instability to overcome Landau damping \cite{Kulsrud69} as well as damping by the background turbulence \cite{Farmer04,Yan04,Lazarian16,Fang19}.  

The magnetic field in ISM generally has a mean direction within one coherent length, which is typically $\sim 50-100\,$pc { \cite{Cho09, Chepurnov10, Beck16}} and comparable to the size of the TeV halo.  
1D particle diffusion actually cannot hold in this scenario, since particles diffuse faster along the mean magnetic field than they diffuse perpendicular to the mean magnetic field in the case of sub-Alfv{\'e}nic turbulence { \citep{Yan08, Giacinti12_prl, Nava13, Lopez18}}. Due to the anisotropy of turbulence in this case, the perpendicular diffusion coefficient is given by $D_{\perp}=D_{\|}M_A^4$ \citep{Yan08, Xu13}, where $D_\|$ is the diffusion coefficient parallel to the magnetic field, $M_A\equiv \delta B_{\rm inj}/B$ is the Alfv{\'e}nic Mach number, which is not far from unity for ISM (i.e., $M_A>0.1$), and $\delta B_{\rm inj}$ is the magnetic perturbation at the injection scale of magnetohydrodynamic (MHD) turbulence or coherence length of magnetic field. Also, the synchrotron radiation intensity becomes anisotropic. Electrons that move along the magnetic field will radiate much less efficiently than those move perpendicular to the magnetic field. Therefore, if the mean magnetic field in the vicinity of Geminga has small inclination toward our line of sight (LOS), the observed synchrotron radiation flux would be much reduced compared to that with the assumption of an isotropic magnetic field, while the diffusion perpendicular to the LOS is slow as suggested by the TeV observation. { Besides, the small inclination is also beneficial to reproduce the isotropic morphology of Geminga's TeV halo \citep{Lopez18}.}

In this letter, we show that both X-ray and TeV observations can be explained with typical conditions for ISM, such as the magnetic field, the diffusion coefficient and the field perturbation level, by considering anisotropic particle diffusion which is a natural outcome in the presence of sub-Alfv{\'e}nic turbulence. We will see that the viewing angle plays an important role in determining the observation signals.

{\it Method-} The temporal evolution of particle number density in space and energy space $N$ is governed by the transport equation 
\begin{equation}
\frac{\partial N}{\partial t}=\nabla \cdot (\bm{\mathcal{D}}\cdot\nabla N)-\frac{\partial}{\partial E_e}\left(\dot{E}_eN\right)+Q
\end{equation}
where $\dot{E}_e$ is the cooling rate of electrons due to synchrotron radiation in ISM magnetic field which is assumed to be $B=3\mu$G in this work, and IC radiation {\bf in the interstellar radiation field with considering the Klein-Nishina effect }. Following \citep{HAWC17_Geminga}, in addition to CMB, we also consider an infrared photon field (with temperature 20\,K and energy density 0.3\,$\rm eVcm^{-3}$), and an optical photon field (with temperature 5000\,K and energy density 0.3\,$\rm eVcm^{-3}$). $Q$ is the source term depicting the electron injection from the pulsar. $\bm{\mathcal{D}}$ is the diffusion tensor. For simplicity, we solve the equation in the cylinder coordinate, defining the $z$-axis to be the direction of the mean magnetic field and the pulsar location to be the origin. By further assuming the system to be symmetric with respect to the $z$-axis (i.e., $\partial/\partial \theta=0$), we can write the transport equation into
\begin{equation}\label{eq:pde}
\begin{split}
\frac{\partial N}{\partial t}=&\frac{1}{r}\frac{\partial}{\partial r}\left(rD_{rr}\frac{\partial N}{\partial r} \right)+D_{zz}\frac{\partial^2N}{\partial z^2}\\
&-\frac{\partial}{\partial E_e}\left(\dot{E}_eN\right)+Q(E_e)S(t)\delta(r)\delta(z).
\end{split}
\end{equation}
where the diffusion coefficient parallel to the mean magnetic field and perpendicular to it can be set, respectively, by
\begin{eqnarray}
D_{zz}=D_{\parallel}=D_0(E_e/{1\rm GeV})^{q}\\
D_{rr}=D_{\perp}=D_{zz}M_A^4
\end{eqnarray}
Here we neglect the drift effect which would cause asymmetric diffusion, and take $D_{\|}$ to be the typical ISM diffusion coefficient throughout the work which is $D_0=3.8\times 10^{28}\rm cm^2s^{-1}$ and $q=1/3$ \cite{Trotta11}. Based on our motivation in this study, we will only look into the case with $M_A=0.1,0.2,0.3$ respectively, since a larger $M_A$ would result in a less anisotropic magnetic field topology. The rightmost term in Eq.~\ref{eq:pde} consists of three parts: the Dirac functions $\delta(r)$ and $\delta(z)$ specify the injection location, $Q(E_e,t)$ represents the injection spectrum of electron, and $S(t)$ shows the temporal behavior of the injection rate. More specifically, we assume the injection spectrum of electron to follow a power-law distribution $Q(E_e)=N_0E_e^{-p}e^{E_e/E_{\rm max}}$, starting from 1\,GeV. Here, $N_0$ is the normalization constant, $p$ is the spectral index and $E_{\rm max}$ is the high-energy cutoff energy in the spectrum. \citep{Xi18} reported a null detection of the diffuse multi-GeV emission from the vicinity of Geminga by \textit{Fermi}-LAT, suggesting a hard injection spectrum of electron from the PWN. We fix the value of $p$ to be 1.6 in this work, noting that the value of $p$ is actually not important to the predicted X-ray flux as long as the TeV observation is reproduced. $E_{\rm max}$ is assumed to be $200\,$TeV to produce a proper spectral shape measured by HAWC under the hard injection spectrum. Assuming the pulsar to be a pure dipole radiator with a braking index of 3, we have $S(t)=(1+t/\tau)^{-2}$ with $\tau=12\,$kyr being the spin-down timescale of the pulsar. The value of $N_0$ is then determined by $\int\int S(t)Q(E_e)dE_edt=W_e$, i.e., the total injected energy in CRe.  
The evolution of the differential electron density $N(E_e, r, z, t)$ is solved by a finite difference method (see Supplement for details).
 
\begin{figure}[htbp]
\centering
\includegraphics[width=0.8\columnwidth]{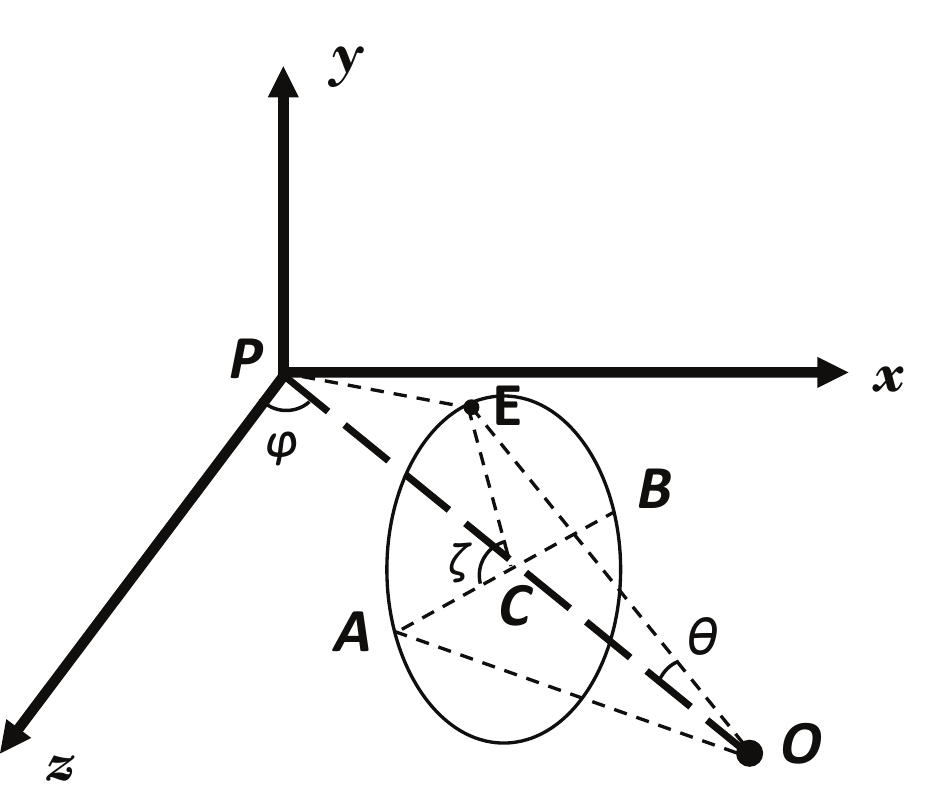}
\caption{Sketch figure for the geometry. See text for more details.}\label{fig:illustration}
\end{figure}

Next, we calculate the emissivity of electrons in the Cartesian coordinate system. Again, we put the pulsar at the origin ($x_P=0, y_P=0, z_P=0$) and define the direction of the mean magnetic field as the $z$-axis. We define the $x-$axis so that the line connecting the pulsar and the observer, i.e., $\overline{PO}$, is in the $xz$ plane (see Fig.~\ref{fig:illustration} for a sketch). Then we envisage a random point $E$ in the space and denote the distance between the point to the observer (i.e., the length of $\overline{EO}$) by $l$, and denote the angle between the line $\overline{EO}$ and the line $\overline{PO}$ by $\theta$. Now let us further consider a circle perpendicular to both the $xz$ plane and the line $\overline{PO}$, with its center, denoted by $C$, attaching to the line $\overline{PO}$ and with the point $E$ on the ring. The two intersection points of the circle and the $xz$ plane are called point $A$ and $B$ respectively, and we call the angle between the line $\overline{AC}$ and the line $\overline{CE}$ angle $\zeta$. The coordinates of point $E$ can then be given by
\begin{eqnarray}
x_E=(d_{\rm gem}-l\cos\theta)\sin\phi - l\sin\theta\cos\zeta\cos\phi \\
y_E=l\sin\theta\sin\zeta \\
z_E=(d_{\rm gem}-l\cos\theta)\cos\phi + l\sin\theta \cos\zeta \sin\phi
\end{eqnarray}

\begin{figure*}[htbp]
\includegraphics[width=1\textwidth]{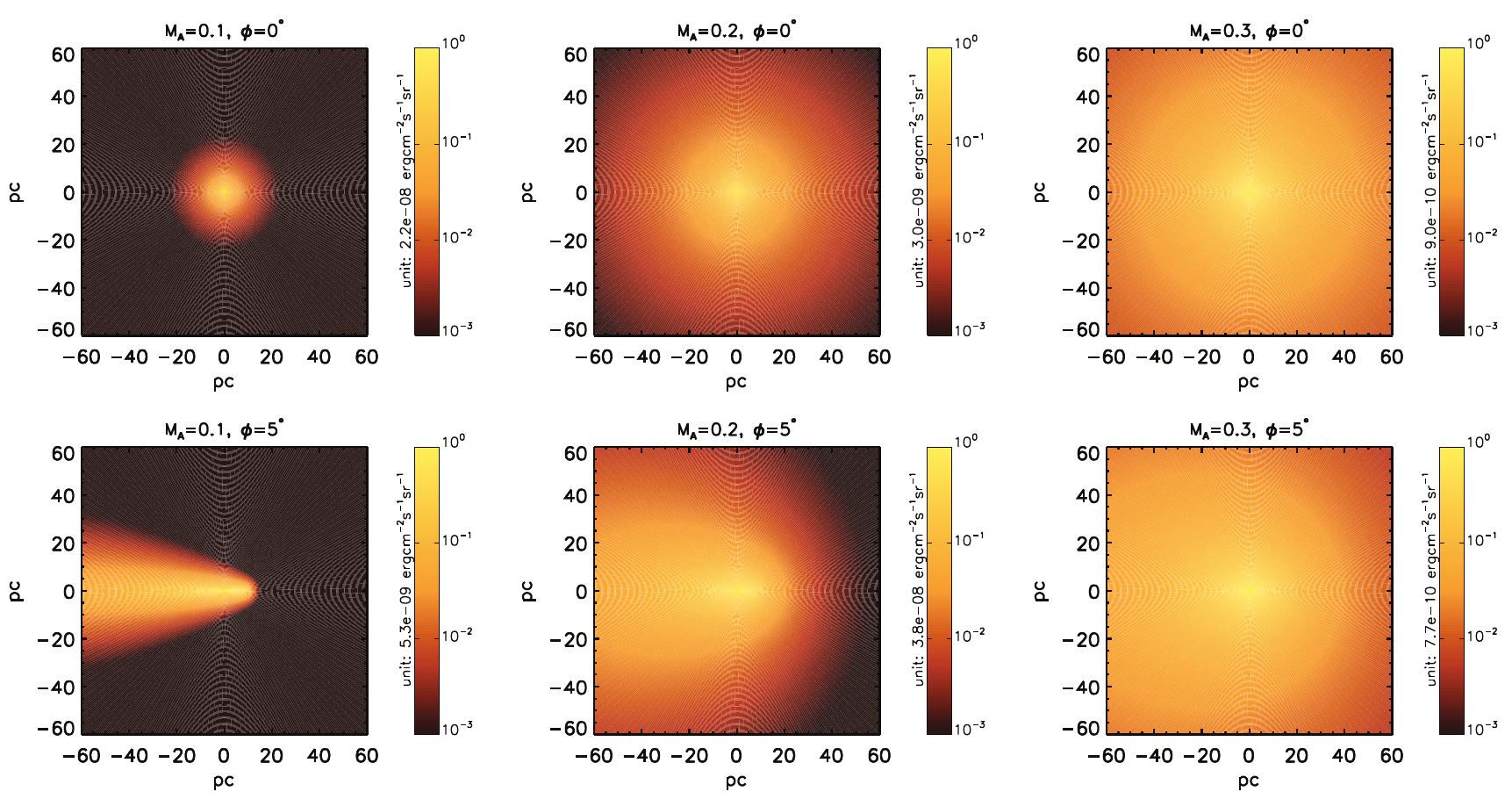}
\caption{Predicted $8-40$\,TeV SBP with different Alfv{\'e}nic Mach number $M_A=0.1, 0.2, 0.3$ and different viewing angle $\phi=0^\circ,5^\circ$.}\label{fig:image}
\end{figure*}

The electron density at an arbitrary point $E$ can be found by interpolation given $z_E$ and $r_E=\sqrt{x_E^2+y_E^2}$, based on the obtained electron density distribution $N(E,z,r)$. The number of electron in the element volume in the neighbourhood of point $E$ can then be given by
\begin{equation}\label{eq:dN}
dN(E,z,r)=N(E,z,r)(l\sin\theta d\zeta \cdot ld\theta \cdot dl)
\end{equation}
where the quantity in the bracket represents the element volume around point $E$. Note that, although the angular distribution of electron is still isotropic (i.e., $dN
/d\Omega=N(E,z,r)/4\pi$ since the mean scattering time of an electron $\sim D/c^2$ is much shorter than the cooling timescale in the energy range of interest in this work), the synchrotron radiation is anisotropic given a mean orientation of magnetic field considered in this work. Due to the relativistic beaming effect, we can only receive the radiation of electrons moving towards us and the radiation power highly depends on the pitch angle $\alpha$ with respect to the local magnetic field. The latter one is determined by the viewing angle $\phi$ and the position of the point $E$. If there is no magnetic field perturbation, the pitch angle can be given by $\cos\alpha_0\equiv \overrightarrow{EO}\cdot \vec{u}_z/\overline{EO}=(z_O-z_E)/l=\cos\theta \cos\phi - \sin\theta \sin\phi \cos\zeta$, where $\vec{u}_z$ is the unit vector along $z$-axis. In the presence of perturbation, the local magnetic field direction will deviate from the $z$-axis by an angle $\delta$. The average cosine of the pitch angle then becomes $\cos\alpha=\cos\alpha_0 \cos\delta$. The distribution of $\cos^2\delta$, i.e., $f(\cos^2\delta)$, where $\int f(\cos^2\delta)d\cos^2\delta=1$, { is obtained from MHD simulations for different $M_A$ \footnote{see Supplemental Material [url] for brief description, which includes Ref.[21-26].}. The average value of $\cos^2\delta$ is close to unity.}  The flux of synchrotron radiation by electrons in the element volume in the neighbourhood of point $E$ can then be given by
\begin{equation}
\begin{split}
dF_{\rm syn}(\epsilon)=&\int \mathcal{F}_{\rm syn}\left\{dN(E,z,r), B\sin\alpha\right\}\\
&\times f(\cos^2\delta)d\cos^2\delta/4\pi l^2
\end{split}
\end{equation}
The IC radiation is isotropic and flux can be given by
\begin{equation}
dF_{\rm IC}(\epsilon)=\mathcal{F}_{\rm IC}\left\{dN(E,z,r), n_{\rm ph}\right\}/4\pi l^2
\end{equation}
where $n_{\rm ph}$ is the differential density of the background photon field.

For observers at Earth, radiation of any electrons in LOS adds up and is projected onto the celestial sphere. The intensity at any given direction depicted by $\theta$ and $\zeta$ can then be found by $I(\epsilon,\theta,\zeta)=\int dF/\sin\theta d\theta d\zeta$. More specifically, the intensity of synchrotron radiation and the intensity of IC radiation can be given by
\begin{equation}
\begin{split}
I_{\rm syn}(\epsilon,\theta, \zeta)=&\frac{1}{4\pi}\int_{\cos^2\delta}\int_{l_{\rm min}}^{l_{\rm max}}\mathcal{F}_{\rm syn}\left\{N(E,z,r),B_0\sin\alpha\right\}\\
&\times f(\cos^2\delta)d\cos^2\delta dl
\end{split}
\end{equation}
and 
\begin{equation}
I_{\rm IC}(\epsilon,\theta,\zeta)=\frac{1}{4\pi}\int_{l_{\rm min}}^{l_{\rm max}}\mathcal{F}_{\rm IC}\left\{N(E,z,r), n_{\rm ph}\right\}dl,
\end{equation}
respectively, where $n_{\rm ph}$ is the photon number density of the background radiation.
The total flux within certain angle $\theta_0$ from the pulsar can be obtained by $F(\epsilon, \theta<\theta_0)=\int_0^{\theta_0}\int_{0}^{2\pi}I(\epsilon,\theta,\zeta)\sin\theta d\theta d\zeta$, where $I=I_{\rm syn}+I_{\rm IC}$.

{\it Result-} We firstly show the predicted 8-40\,TeV gamma-ray morphology for different Alfv{\'e}nic Mach number $M_A$ and different viewing angle $\phi$ in Fig.~\ref{fig:image}. The Geminga pulsar is located at the center of each panel or the coordinate (0,0). The horizontal axis is parallel to the line $\overline{AB}$ while the vertical axis is parallel to $y-$axis in Fig.~\ref{fig:illustration}. The projected distance is calculated based on a nominal distance of 250\,pc for Geminga. We can see that the morphology is too compact in the case of $\phi=0^\circ, M_A=0.1$. This is because the perpendicular diffusion coefficient is only $D_{\perp}=3.8\times 10^{24}(E_e/1\rm GeV)^{1/3}\rm cm^2s^{-1}$ for $M_A=0.1$, and the perpendicular diffusion distance is correspondingly only $\sim 5\,$pc within the TeV-emitting electron's cooling timescale which is $t_c\lesssim 10^{12}$s. For a viewing angle of $\phi=0^\circ$, such a perpendicular diffusion length is translated to only $\sim 1^\circ$ extension in the celestial sphere. The morphology is highly anisotropic in the case of $\phi=5^\circ, M_A=0.1$ which is obviously inconsistent with the observation. { This is because the LOS towards the left side of the pulsar (e.g. direction of $\overline{OA}$) passes through more electrons than the LOS towards the right side of the pulsar (e.g., direction of $\overline{OB}$).} On the other hand, in both two cases with $M_A=0.3$, the morphology does not show a sufficient gradient as that observed by HAWC. We therefore focus on the cases of $\phi=0^\circ$ and $\phi=5^\circ$ with $M_A=0.2$ below. {\bf A larger $\phi$ would result in a more anisotropic morphology and a higher X-ray flux so we do not consider it here.} 

\begin{figure}[htbp]
\centering
\includegraphics[width=1\columnwidth]{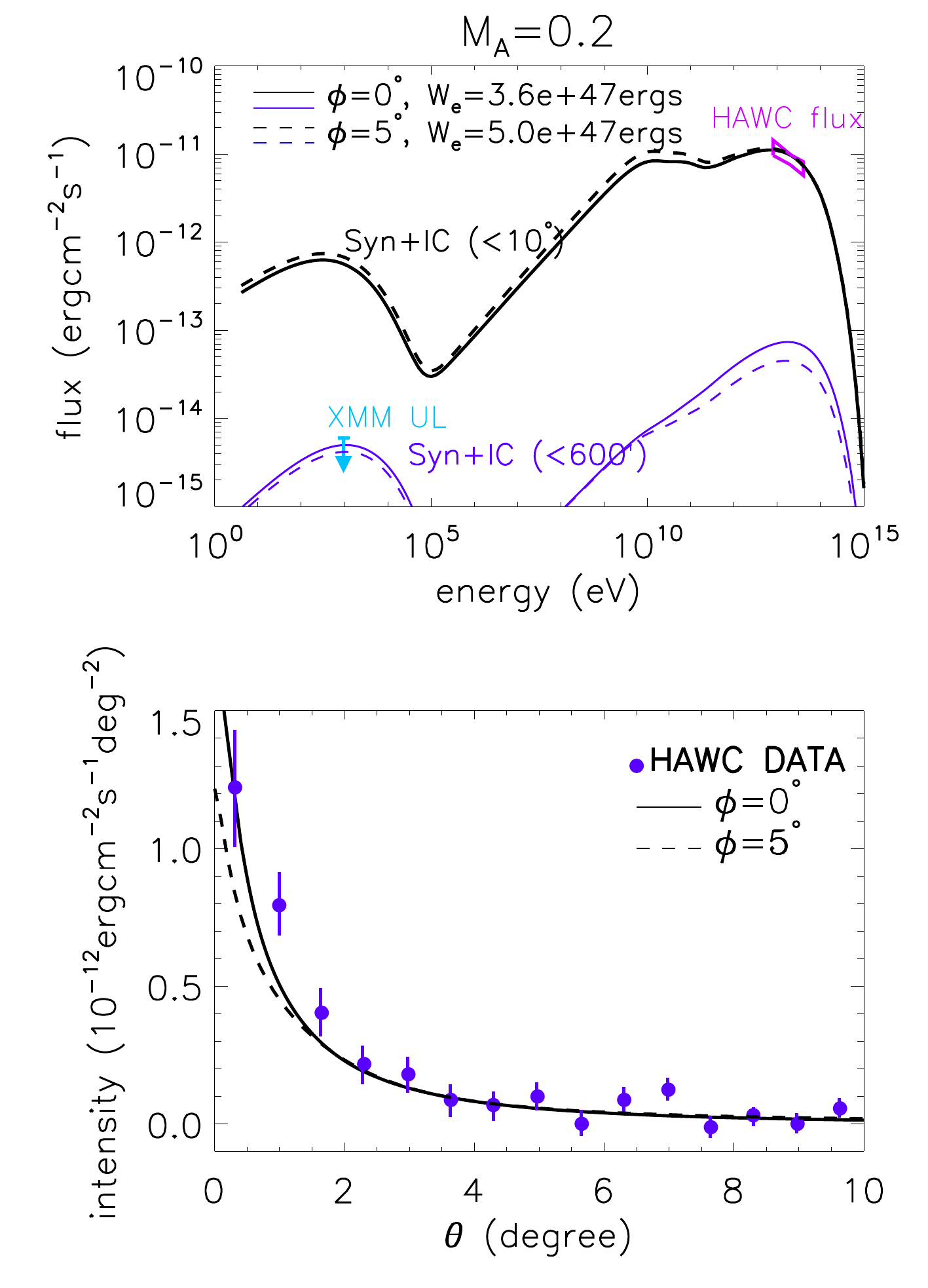}
\caption{Results with $\phi=0^\circ$ and $\phi=5^\circ$ for $M_A=0.2$. {\bf Upper:} the predicted multiwavelength flux from a region within $10^\circ$ from Geminga (black curves) and from a region within $600''$ from Geminga (blue curves). Solid curves represent the result of $\phi=0^\circ$ while dashed curves represent the result of $\phi=5^\circ$. The magenta bowtie and the cyan arrow represent the flux measured by HAWC and the upper limit from XMM-Newton respectively. {\bf Lower:} the predicted 1D ($\zeta$-averaged) SBP in $8-40$\,TeV in comparison with the measured one by HAWC, which is shown as blue circles.}\label{fig:result}
\end{figure}

We integrate the TeV emission and X-ray emission over a circular region with a radius of $10^\circ$ and $600''$ centred at Geminga, respectively, to compare with predicted fluxers with HAWC's observation and XMM-Newton upper limit. In the calculation, { we adjust the value of parameter $W_e$ to normalize the predicted TeV flux to the measured one}. As we can see from Fig.~\ref{fig:result}, the predicted X-ray fluxes are lower than the upper limit of XMM-Newton in both two cases. The predicted surface brightness profile (SBP) in $8-40\,$TeV is in good agreement with the observation for $\phi=0^\circ$. For $\phi=5^\circ$, the predicted SBP is a little flatter than the observation. {\bf The reduced Chi-square test returns $\chi^2/dof=1.73$ with $dof=14$ being the degrees of freedom in the fitting.  It corresponds to a $p$-value of 0.044, implying the fitting is marginally consistent with the data at $2\sigma$ level. We therefore conclude that there should be a magnetic field alignment $\lesssim 5^\circ$ with our LOS to explain the observation for $M_A\simeq 0.2$.} The small inclination between the mean magnetic field and LOS is consistent with the synchrotron polarization measurement by \cite{Gao10} on the region of $\sim 1^\circ$ around Geminga showing very small plane-of-sky magnetic field component.  Our model can be tested in the future after many such TeV halos being detected with well determined morphologies, since it would expect (in a flux-limited sample) to observe a large number of elongated systems ($B$ not aligned along LOS) and a minority of roughly spherical ones ($B$ aligned along LOS) provided that $M_A$ being significantly smaller than unity. {\bf For a larger $M_A$, the particle diffusion becomes more isotropic and consequently the morphology of the TeV halo would be less dependent on $\phi$.}


Lastly, we note that given the distance from Geminga to Earth being 250\,pc, the intervening ISM is expected to contain two or three coherent magnetic fields. The mean magnetic field is unlikely always aligned with our LOS between Geminga and Earth. The pitch angle between the electron moving towards us and the mean field is supposed to be larger outside the TeV halo. The potential increase of synchrotron radiation flux is, however, limited \footnote{Even in the extreme case of a mean magnetic field perpendicular to LOS (or $90^\circ$ pitch angle) outside the TeV halo, the synchrotron flux can still be consistent with the upper limit by employing a slightly weaker magnetic field of $2\mu$G for the intervening medium.}. Another related issue is the contribution of Geminga to the positron excess measured in many experiments above 10\,GeV \citep{Adriani09, Fermi12, Aguilar14_ee+}. There is a debate on the contribution of Geminga to the positron excess in the framework of inefficient (isotropic) diffusion of particles\citep{Hooper17,Fang18,Profumo18,Xi18,Tang19, Dimauro19}. 
In the global frame from Geminga to Earth, the injected positrons from Geminga are likely to diffuse fast with the typical ISM diffusion coefficient so that the resulting positron flux at Earth {\bf can be enhanced compared to the isotropic diffusion scenario, whereas the exact flux will depend on properties of turbulence between Earth and Geminga.} 



\bibliography{ms.bib}

\begin{thebibliography}{34}
\expandafter\ifx\csname natexlab\endcsname\relax\def\natexlab#1{#1}\fi
\expandafter\ifx\csname bibnamefont\endcsname\relax
  \def\bibnamefont#1{#1}\fi
\expandafter\ifx\csname bibfnamefont\endcsname\relax
  \def\bibfnamefont#1{#1}\fi
\expandafter\ifx\csname citenamefont\endcsname\relax
  \def\citenamefont#1{#1}\fi
\expandafter\ifx\csname url\endcsname\relax
  \def\url#1{\texttt{#1}}\fi
\expandafter\ifx\csname urlprefix\endcsname\relax\def\urlprefix{URL }\fi
\providecommand{\bibinfo}[2]{#2}
\providecommand{\eprint}[2][]{\url{#2}}

\bibitem[{\citenamefont{{Abeysekara} et~al.}(2017)\citenamefont{{Abeysekara},
  {Albert}, {Alfaro}, {Alvarez}, {{\'A}lvarez}, and et~al.}}]{HAWC17_Geminga}
\bibinfo{author}{\bibfnamefont{A.~U.} \bibnamefont{{Abeysekara}}},
  \bibinfo{author}{\bibfnamefont{A.}~\bibnamefont{{Albert}}},
  \bibinfo{author}{\bibfnamefont{R.}~\bibnamefont{{Alfaro}}},
  \bibinfo{author}{\bibfnamefont{C.}~\bibnamefont{{Alvarez}}},
  \bibinfo{author}{\bibfnamefont{J.~D.} \bibnamefont{{{\'A}lvarez}}},
  \bibnamefont{and} \bibinfo{author}{\bibnamefont{et~al.}},
  \bibinfo{journal}{Science} \textbf{\bibinfo{volume}{358}},
  \bibinfo{pages}{911} (\bibinfo{year}{2017}), \eprint{1711.06223}.

\bibitem[{\citenamefont{{Fang} et~al.}(2018)\citenamefont{{Fang}, {Bi}, {Yin},
  and {Yuan}}}]{Fang18}
\bibinfo{author}{\bibfnamefont{K.}~\bibnamefont{{Fang}}},
  \bibinfo{author}{\bibfnamefont{X.-J.} \bibnamefont{{Bi}}},
  \bibinfo{author}{\bibfnamefont{P.-F.} \bibnamefont{{Yin}}}, \bibnamefont{and}
  \bibinfo{author}{\bibfnamefont{Q.}~\bibnamefont{{Yuan}}},
  \bibinfo{journal}{\apj} \textbf{\bibinfo{volume}{863}}, \bibinfo{eid}{30}
  (\bibinfo{year}{2018}), \eprint{1803.02640}.

\bibitem[{\citenamefont{{Profumo} et~al.}(2018)\citenamefont{{Profumo},
  {Reynoso-Cordova}, {Kaaz}, and {Silverman}}}]{Profumo18}
\bibinfo{author}{\bibfnamefont{S.}~\bibnamefont{{Profumo}}},
  \bibinfo{author}{\bibfnamefont{J.}~\bibnamefont{{Reynoso-Cordova}}},
  \bibinfo{author}{\bibfnamefont{N.}~\bibnamefont{{Kaaz}}}, \bibnamefont{and}
  \bibinfo{author}{\bibfnamefont{M.}~\bibnamefont{{Silverman}}},
  \bibinfo{journal}{\prd} \textbf{\bibinfo{volume}{97}}, \bibinfo{eid}{123008}
  (\bibinfo{year}{2018}), \eprint{1803.09731}.

\bibitem[{\citenamefont{{Xi} et~al.}(2018)\citenamefont{{Xi}, {Liu}, {Huang},
  {Fang}, {Yan}, and {Wang}}}]{Xi18}
\bibinfo{author}{\bibfnamefont{S.-Q.} \bibnamefont{{Xi}}},
  \bibinfo{author}{\bibfnamefont{R.-Y.} \bibnamefont{{Liu}}},
  \bibinfo{author}{\bibfnamefont{Z.-Q.} \bibnamefont{{Huang}}},
  \bibinfo{author}{\bibfnamefont{K.}~\bibnamefont{{Fang}}},
  \bibinfo{author}{\bibfnamefont{H.}~\bibnamefont{{Yan}}}, \bibnamefont{and}
  \bibinfo{author}{\bibfnamefont{X.-Y.} \bibnamefont{{Wang}}},
  \bibinfo{journal}{ArXiv e-prints}  (\bibinfo{year}{2018}),
  \eprint{1810.10928}.

\bibitem[{\citenamefont{{Liu} et~al.}(2019)\citenamefont{{Liu}, {Ge}, {Sun},
  and {Wang}}}]{Liu19}
\bibinfo{author}{\bibfnamefont{R.-Y.} \bibnamefont{{Liu}}},
  \bibinfo{author}{\bibfnamefont{C.}~\bibnamefont{{Ge}}},
  \bibinfo{author}{\bibfnamefont{X.-N.} \bibnamefont{{Sun}}}, \bibnamefont{and}
  \bibinfo{author}{\bibfnamefont{X.-Y.} \bibnamefont{{Wang}}},
  \bibinfo{journal}{\apj} \textbf{\bibinfo{volume}{875}}, \bibinfo{eid}{149}
  (\bibinfo{year}{2019}), \eprint{1904.11438}.

\bibitem[{\citenamefont{{Kulsrud} and {Pearce}}(1969)}]{Kulsrud69}
\bibinfo{author}{\bibfnamefont{R.}~\bibnamefont{{Kulsrud}}} \bibnamefont{and}
  \bibinfo{author}{\bibfnamefont{W.~P.} \bibnamefont{{Pearce}}},
  \bibinfo{journal}{\apj} \textbf{\bibinfo{volume}{156}}, \bibinfo{pages}{445}
  (\bibinfo{year}{1969}).

\bibitem[{\citenamefont{{Farmer} and {Goldreich}}(2004)}]{Farmer04}
\bibinfo{author}{\bibfnamefont{A.~J.} \bibnamefont{{Farmer}}} \bibnamefont{and}
  \bibinfo{author}{\bibfnamefont{P.}~\bibnamefont{{Goldreich}}},
  \bibinfo{journal}{\apj} \textbf{\bibinfo{volume}{604}}, \bibinfo{pages}{671}
  (\bibinfo{year}{2004}), \eprint{astro-ph/0311400}.

\bibitem[{\citenamefont{{Yan} and {Lazarian}}(2004)}]{Yan04}
\bibinfo{author}{\bibfnamefont{H.}~\bibnamefont{{Yan}}} \bibnamefont{and}
  \bibinfo{author}{\bibfnamefont{A.}~\bibnamefont{{Lazarian}}},
  \bibinfo{journal}{\apj} \textbf{\bibinfo{volume}{614}}, \bibinfo{pages}{757}
  (\bibinfo{year}{2004}), \eprint{astro-ph/0408172}.

\bibitem[{\citenamefont{{Lazarian}}(2016)}]{Lazarian16}
\bibinfo{author}{\bibfnamefont{A.}~\bibnamefont{{Lazarian}}},
  \bibinfo{journal}{\apj} \textbf{\bibinfo{volume}{833}}, \bibinfo{eid}{131}
  (\bibinfo{year}{2016}), \eprint{1607.02042}.

\bibitem[{\citenamefont{{Fang} et~al.}(2019)\citenamefont{{Fang}, {Bi}, and
  {Yin}}}]{Fang19}
\bibinfo{author}{\bibfnamefont{K.}~\bibnamefont{{Fang}}},
  \bibinfo{author}{\bibfnamefont{X.-J.} \bibnamefont{{Bi}}}, \bibnamefont{and}
  \bibinfo{author}{\bibfnamefont{P.-F.} \bibnamefont{{Yin}}},
  \bibinfo{journal}{arXiv e-prints}  (\bibinfo{year}{2019}),
  \eprint{1903.06421}.

\bibitem[{\citenamefont{{Cho} and {Ryu}}(2009)}]{Cho09}
\bibinfo{author}{\bibfnamefont{J.}~\bibnamefont{{Cho}}} \bibnamefont{and}
  \bibinfo{author}{\bibfnamefont{D.}~\bibnamefont{{Ryu}}},
  \bibinfo{journal}{\apjl} \textbf{\bibinfo{volume}{705}}, \bibinfo{pages}{L90}
  (\bibinfo{year}{2009}), \eprint{0908.0610}.

\bibitem[{\citenamefont{{Chepurnov} and {Lazarian}}(2010)}]{Chepurnov10}
\bibinfo{author}{\bibfnamefont{A.}~\bibnamefont{{Chepurnov}}} \bibnamefont{and}
  \bibinfo{author}{\bibfnamefont{A.}~\bibnamefont{{Lazarian}}},
  \bibinfo{journal}{\apj} \textbf{\bibinfo{volume}{710}}, \bibinfo{pages}{853}
  (\bibinfo{year}{2010}), \eprint{0905.4413}.

\bibitem[{\citenamefont{{Beck} et~al.}(2016)\citenamefont{{Beck}, {Beck},
  {Beck}, {Dolag}, {Strong}, and {Nielaba}}}]{Beck16}
\bibinfo{author}{\bibfnamefont{M.~C.} \bibnamefont{{Beck}}},
  \bibinfo{author}{\bibfnamefont{A.~M.} \bibnamefont{{Beck}}},
  \bibinfo{author}{\bibfnamefont{R.}~\bibnamefont{{Beck}}},
  \bibinfo{author}{\bibfnamefont{K.}~\bibnamefont{{Dolag}}},
  \bibinfo{author}{\bibfnamefont{A.~W.} \bibnamefont{{Strong}}},
  \bibnamefont{and}
  \bibinfo{author}{\bibfnamefont{P.}~\bibnamefont{{Nielaba}}},
  \bibinfo{journal}{\jcap} \textbf{\bibinfo{volume}{2016}}, \bibinfo{eid}{056}
  (\bibinfo{year}{2016}), \eprint{1409.5120}.

\bibitem[{\citenamefont{{Yan} and {Lazarian}}(2008)}]{Yan08}
\bibinfo{author}{\bibfnamefont{H.}~\bibnamefont{{Yan}}} \bibnamefont{and}
  \bibinfo{author}{\bibfnamefont{A.}~\bibnamefont{{Lazarian}}},
  \bibinfo{journal}{\apj} \textbf{\bibinfo{volume}{673}},
  \bibinfo{eid}{942-953} (\bibinfo{year}{2008}), \eprint{0710.2617}.

\bibitem[{\citenamefont{{Giacinti} and {Sigl}}(2012)}]{Giacinti12_prl}
\bibinfo{author}{\bibfnamefont{G.}~\bibnamefont{{Giacinti}}} \bibnamefont{and}
  \bibinfo{author}{\bibfnamefont{G.}~\bibnamefont{{Sigl}}},
  \bibinfo{journal}{\prl} \textbf{\bibinfo{volume}{109}}, \bibinfo{eid}{071101}
  (\bibinfo{year}{2012}), \eprint{1111.2536}.

\bibitem[{\citenamefont{{Nava} and {Gabici}}(2013)}]{Nava13}
\bibinfo{author}{\bibfnamefont{L.}~\bibnamefont{{Nava}}} \bibnamefont{and}
  \bibinfo{author}{\bibfnamefont{S.}~\bibnamefont{{Gabici}}},
  \bibinfo{journal}{\mnras} \textbf{\bibinfo{volume}{429}},
  \bibinfo{pages}{1643} (\bibinfo{year}{2013}), \eprint{1211.1668}.

\bibitem[{\citenamefont{{L{\'o}pez-Coto} and {Giacinti}}(2018)}]{Lopez18}
\bibinfo{author}{\bibfnamefont{R.}~\bibnamefont{{L{\'o}pez-Coto}}}
  \bibnamefont{and}
  \bibinfo{author}{\bibfnamefont{G.}~\bibnamefont{{Giacinti}}},
  \bibinfo{journal}{\mnras} \textbf{\bibinfo{volume}{479}},
  \bibinfo{pages}{4526} (\bibinfo{year}{2018}), \eprint{1712.04373}.

\bibitem[{\citenamefont{{Xu} and {Yan}}(2013)}]{Xu13}
\bibinfo{author}{\bibfnamefont{S.}~\bibnamefont{{Xu}}} \bibnamefont{and}
  \bibinfo{author}{\bibfnamefont{H.}~\bibnamefont{{Yan}}},
  \bibinfo{journal}{\apj} \textbf{\bibinfo{volume}{779}}, \bibinfo{eid}{140}
  (\bibinfo{year}{2013}), \eprint{1307.1346}.

\bibitem[{\citenamefont{{Trotta} et~al.}(2011)\citenamefont{{Trotta},
  {J{\'o}hannesson}, {Moskalenko}, {Porter}, {Ruiz de Austri}, and
  {Strong}}}]{Trotta11}
\bibinfo{author}{\bibfnamefont{R.}~\bibnamefont{{Trotta}}},
  \bibinfo{author}{\bibfnamefont{G.}~\bibnamefont{{J{\'o}hannesson}}},
  \bibinfo{author}{\bibfnamefont{I.~V.} \bibnamefont{{Moskalenko}}},
  \bibinfo{author}{\bibfnamefont{T.~A.} \bibnamefont{{Porter}}},
  \bibinfo{author}{\bibfnamefont{R.}~\bibnamefont{{Ruiz de Austri}}},
  \bibnamefont{and} \bibinfo{author}{\bibfnamefont{A.~W.}
  \bibnamefont{{Strong}}}, \bibinfo{journal}{\apj}
  \textbf{\bibinfo{volume}{729}}, \bibinfo{eid}{106} (\bibinfo{year}{2011}),
  \eprint{1011.0037}.

\bibitem[{Note1()}]{Note1}
\bibinfo{note}{see Supplemental Material [url] for brief description,
  which includes Ref.[21-26].}

\bibitem[{\citenamefont{{Makwana} and {Yan}}(2019)}]{Makwana19}
\bibinfo{author}{\bibfnamefont{K.~D.} \bibnamefont{{Makwana}}}
  \bibnamefont{and} \bibinfo{author}{\bibfnamefont{H.}~\bibnamefont{{Yan}}},
  \bibinfo{journal}{arXiv e-prints} \bibinfo{eid}{arXiv:1907.01853}
  (\bibinfo{year}{2019}).

\bibitem[{\citenamefont{{Brandenburg} and {Dobler}}(2002)}]{Brandenburg02}
\bibinfo{author}{\bibfnamefont{A.}~\bibnamefont{{Brandenburg}}}
  \bibnamefont{and} \bibinfo{author}{\bibfnamefont{W.}~\bibnamefont{{Dobler}}},
  \bibinfo{journal}{Computer Physics Communications}
  \textbf{\bibinfo{volume}{147}}, \bibinfo{pages}{471} (\bibinfo{year}{2002}),
  \eprint{astro-ph/0111569}.

\bibitem[{\citenamefont{{Brandenburg} and {Dobler}}(2010)}]{Pencil10}
\bibinfo{author}{\bibfnamefont{A.}~\bibnamefont{{Brandenburg}}}
  \bibnamefont{and} \bibinfo{author}{\bibfnamefont{W.}~\bibnamefont{{Dobler}}},
  \emph{\bibinfo{title}{{Pencil: Finite-difference Code for Compressible
  Hydrodynamic Flows}}} (\bibinfo{year}{2010}), \eprint{1010.060}.

\bibitem[{\citenamefont{{Mignone} et~al.}(2007)\citenamefont{{Mignone}, {Bodo},
  {Massaglia}, {Matsakos}, {Tesileanu}, {Zanni}, and {Ferrari}}}]{Pluto07}
\bibinfo{author}{\bibfnamefont{A.}~\bibnamefont{{Mignone}}},
  \bibinfo{author}{\bibfnamefont{G.}~\bibnamefont{{Bodo}}},
  \bibinfo{author}{\bibfnamefont{S.}~\bibnamefont{{Massaglia}}},
  \bibinfo{author}{\bibfnamefont{T.}~\bibnamefont{{Matsakos}}},
  \bibinfo{author}{\bibfnamefont{O.}~\bibnamefont{{Tesileanu}}},
  \bibinfo{author}{\bibfnamefont{C.}~\bibnamefont{{Zanni}}}, \bibnamefont{and}
  \bibinfo{author}{\bibfnamefont{A.}~\bibnamefont{{Ferrari}}},
  \bibinfo{journal}{\apjs} \textbf{\bibinfo{volume}{170}}, \bibinfo{pages}{228}
  (\bibinfo{year}{2007}), \eprint{astro-ph/0701854}.

\bibitem[{\citenamefont{{Mignone} et~al.}(2012)\citenamefont{{Mignone},
  {Zanni}, {Tzeferacos}, {van Straalen}, {Colella}, and {Bodo}}}]{Pluto12}
\bibinfo{author}{\bibfnamefont{A.}~\bibnamefont{{Mignone}}},
  \bibinfo{author}{\bibfnamefont{C.}~\bibnamefont{{Zanni}}},
  \bibinfo{author}{\bibfnamefont{P.}~\bibnamefont{{Tzeferacos}}},
  \bibinfo{author}{\bibfnamefont{B.}~\bibnamefont{{van Straalen}}},
  \bibinfo{author}{\bibfnamefont{P.}~\bibnamefont{{Colella}}},
  \bibnamefont{and} \bibinfo{author}{\bibfnamefont{G.}~\bibnamefont{{Bodo}}},
  \bibinfo{journal}{\apjs} \textbf{\bibinfo{volume}{198}}, \bibinfo{eid}{7}
  (\bibinfo{year}{2012}), \eprint{1110.0740}.

\bibitem[{\citenamefont{{Lazarian} and {Pogosyan}}(2012)}]{Lazarian12}
\bibinfo{author}{\bibfnamefont{A.}~\bibnamefont{{Lazarian}}} \bibnamefont{and}
  \bibinfo{author}{\bibfnamefont{D.}~\bibnamefont{{Pogosyan}}},
  \bibinfo{journal}{\apj} \textbf{\bibinfo{volume}{747}}, \bibinfo{eid}{5}
  (\bibinfo{year}{2012}), \eprint{1105.4617}.

\bibitem[{\citenamefont{{Gao} et~al.}(2010)\citenamefont{{Gao}, {Reich}, {Han},
  {Sun}, {Wielebinski}, {Shi}, {Xiao}, {Reich}, {F{\"u}rst}, {Chen}
  et~al.}}]{Gao10}
\bibinfo{author}{\bibfnamefont{X.~Y.} \bibnamefont{{Gao}}},
  \bibinfo{author}{\bibfnamefont{W.}~\bibnamefont{{Reich}}},
  \bibinfo{author}{\bibfnamefont{J.~L.} \bibnamefont{{Han}}},
  \bibinfo{author}{\bibfnamefont{X.~H.} \bibnamefont{{Sun}}},
  \bibinfo{author}{\bibfnamefont{R.}~\bibnamefont{{Wielebinski}}},
  \bibinfo{author}{\bibfnamefont{W.~B.} \bibnamefont{{Shi}}},
  \bibinfo{author}{\bibfnamefont{L.}~\bibnamefont{{Xiao}}},
  \bibinfo{author}{\bibfnamefont{P.}~\bibnamefont{{Reich}}},
  \bibinfo{author}{\bibfnamefont{E.}~\bibnamefont{{F{\"u}rst}}},
  \bibinfo{author}{\bibfnamefont{M.~Z.} \bibnamefont{{Chen}}},
  \bibnamefont{et~al.}, \bibinfo{journal}{\aap} \textbf{\bibinfo{volume}{515}},
  \bibinfo{eid}{A64} (\bibinfo{year}{2010}), \eprint{1004.4072}.

\bibitem[{Note2()}]{Note2}
\bibinfo{note}{even in the extreme case of a mean magnetic field
  perpendicular to LOS (or $90^\circ $ pitch angle) outside the TeV halo, the
  synchrotron flux can still be consistent with the upper limit by employing a
  slightly weaker magnetic field of $2\mu $G for the intervening medium.}

\bibitem[{\citenamefont{{Adriani} et~al.}(2009)\citenamefont{{Adriani},
  {Barbarino}, {Bazilevskaya}, {Bellotti}, {Boezio}, {Bogomolov}, {Bonechi},
  {Bongi}, {Bonvicini}, {Bottai} et~al.}}]{Adriani09}
\bibinfo{author}{\bibfnamefont{O.}~\bibnamefont{{Adriani}}},
  \bibinfo{author}{\bibfnamefont{G.~C.} \bibnamefont{{Barbarino}}},
  \bibinfo{author}{\bibfnamefont{G.~A.} \bibnamefont{{Bazilevskaya}}},
  \bibinfo{author}{\bibfnamefont{R.}~\bibnamefont{{Bellotti}}},
  \bibinfo{author}{\bibfnamefont{M.}~\bibnamefont{{Boezio}}},
  \bibinfo{author}{\bibfnamefont{E.~A.} \bibnamefont{{Bogomolov}}},
  \bibinfo{author}{\bibfnamefont{L.}~\bibnamefont{{Bonechi}}},
  \bibinfo{author}{\bibfnamefont{M.}~\bibnamefont{{Bongi}}},
  \bibinfo{author}{\bibfnamefont{V.}~\bibnamefont{{Bonvicini}}},
  \bibinfo{author}{\bibfnamefont{S.}~\bibnamefont{{Bottai}}},
  \bibnamefont{et~al.}, \bibinfo{journal}{\nat} \textbf{\bibinfo{volume}{458}},
  \bibinfo{pages}{607} (\bibinfo{year}{2009}), \eprint{0810.4995}.

\bibitem[{\citenamefont{{Ackermann} et~al.}(2012)\citenamefont{{Ackermann},
  {Ajello}, {Allafort}, {Atwood}, {Baldini}, {Barbiellini}, {Bastieri},
  {Bechtol}, {Bellazzini}, {Berenji} et~al.}}]{Fermi12}
\bibinfo{author}{\bibfnamefont{M.}~\bibnamefont{{Ackermann}}},
  \bibinfo{author}{\bibfnamefont{M.}~\bibnamefont{{Ajello}}},
  \bibinfo{author}{\bibfnamefont{A.}~\bibnamefont{{Allafort}}},
  \bibinfo{author}{\bibfnamefont{W.~B.} \bibnamefont{{Atwood}}},
  \bibinfo{author}{\bibfnamefont{L.}~\bibnamefont{{Baldini}}},
  \bibinfo{author}{\bibfnamefont{G.}~\bibnamefont{{Barbiellini}}},
  \bibinfo{author}{\bibfnamefont{D.}~\bibnamefont{{Bastieri}}},
  \bibinfo{author}{\bibfnamefont{K.}~\bibnamefont{{Bechtol}}},
  \bibinfo{author}{\bibfnamefont{R.}~\bibnamefont{{Bellazzini}}},
  \bibinfo{author}{\bibfnamefont{B.}~\bibnamefont{{Berenji}}},
  \bibnamefont{et~al.}, \bibinfo{journal}{Physical Review Letters}
  \textbf{\bibinfo{volume}{108}}, \bibinfo{eid}{011103} (\bibinfo{year}{2012}),
  \eprint{1109.0521}.

\bibitem[{\citenamefont{{Aguilar} et~al.}(2014)\citenamefont{{Aguilar}, {Aisa},
  {Alvino}, {Ambrosi}, {Andeen}, {Arruda}, {Attig}, {Azzarello}, {Bachlechner},
  {Barao} et~al.}}]{Aguilar14_ee+}
\bibinfo{author}{\bibfnamefont{M.}~\bibnamefont{{Aguilar}}},
  \bibinfo{author}{\bibfnamefont{D.}~\bibnamefont{{Aisa}}},
  \bibinfo{author}{\bibfnamefont{A.}~\bibnamefont{{Alvino}}},
  \bibinfo{author}{\bibfnamefont{G.}~\bibnamefont{{Ambrosi}}},
  \bibinfo{author}{\bibfnamefont{K.}~\bibnamefont{{Andeen}}},
  \bibinfo{author}{\bibfnamefont{L.}~\bibnamefont{{Arruda}}},
  \bibinfo{author}{\bibfnamefont{N.}~\bibnamefont{{Attig}}},
  \bibinfo{author}{\bibfnamefont{P.}~\bibnamefont{{Azzarello}}},
  \bibinfo{author}{\bibfnamefont{A.}~\bibnamefont{{Bachlechner}}},
  \bibinfo{author}{\bibfnamefont{F.}~\bibnamefont{{Barao}}},
  \bibnamefont{et~al.}, \bibinfo{journal}{Physical Review Letters}
  \textbf{\bibinfo{volume}{113}}, \bibinfo{eid}{121102} (\bibinfo{year}{2014}).

\bibitem[{\citenamefont{{Hooper} et~al.}(2017)\citenamefont{{Hooper}, {Cholis},
  {Linden}, and {Fang}}}]{Hooper17}
\bibinfo{author}{\bibfnamefont{D.}~\bibnamefont{{Hooper}}},
  \bibinfo{author}{\bibfnamefont{I.}~\bibnamefont{{Cholis}}},
  \bibinfo{author}{\bibfnamefont{T.}~\bibnamefont{{Linden}}}, \bibnamefont{and}
  \bibinfo{author}{\bibfnamefont{K.}~\bibnamefont{{Fang}}},
  \bibinfo{journal}{\prd} \textbf{\bibinfo{volume}{96}}, \bibinfo{eid}{103013}
  (\bibinfo{year}{2017}), \eprint{1702.08436}.

\bibitem[{\citenamefont{{Tang} and {Piran}}(2019)}]{Tang19}
\bibinfo{author}{\bibfnamefont{X.}~\bibnamefont{{Tang}}} \bibnamefont{and}
  \bibinfo{author}{\bibfnamefont{T.}~\bibnamefont{{Piran}}},
  \bibinfo{journal}{\mnras} \textbf{\bibinfo{volume}{484}},
  \bibinfo{pages}{3491} (\bibinfo{year}{2019}), \eprint{1808.02445}.

\bibitem[{\citenamefont{{Di Mauro} et~al.}(2019)\citenamefont{{Di Mauro},
  {Manconi}, and {Donato}}}]{Dimauro19}
\bibinfo{author}{\bibfnamefont{M.}~\bibnamefont{{Di Mauro}}},
  \bibinfo{author}{\bibfnamefont{S.}~\bibnamefont{{Manconi}}},
  \bibnamefont{and} \bibinfo{author}{\bibfnamefont{F.}~\bibnamefont{{Donato}}},
  \bibinfo{journal}{arXiv e-prints} \bibinfo{eid}{arXiv:1903.05647}
  (\bibinfo{year}{2019}), \eprint{1903.05647}.

\end{thebibliography}

\clearpage
\appendix
\begin{widetext}

\section*{Solving the transport equation}
We employ the operator splitting technique to solve the transport equation
\begin{equation}\label{eq:pde}
\frac{\partial N}{\partial t}=\frac{1}{r}\frac{\partial}{\partial r}\left(rD_{rr}\frac{\partial N}{\partial r} \right)+D_{zz}\frac{\partial^2N}{\partial z^2}-\frac{\partial}{\partial E_e}\left(\dot{E}_eN\right)+Q(E_e)S(t)\delta(r)\delta(z),
\end{equation}
 such that the problem can be simplified into solving a convection equation in energy space and a diffusion equation in real space separately. The Strang splitting scheme is used to decouple the energy operator and the spatial operator, with the flow chart as
\begin{equation}
N^l\xrightarrow[(t_l,t_{l+1/2})]{\mathcal{L}_E} \widetilde{N}^{l+1/2}\xrightarrow[(t_l,t_{l+1})]{\mathcal{L}_r} N^{l+1/2}\xrightarrow[(t_{l+1/2},t_{l+1})]{\mathcal{L}_E}N^{l+1},
\end{equation}
with $l$ being the index of the time step. The spatial operator $\mathcal{L}_r$ contain both the $r$-derivative terms and $z$-derivative terms. We further employ the alternating-direction implicit method to divide each time step into two steps of size $\Delta t/2$, and each step can be solved using tridiagonal matrix algorithm.  

The implicit second-order upwind scheme is used to discretize the convection equation in energy space as
\begin{equation}
\begin{split}
\frac{N_{i,j,k}^{l+1}-N_{i,j,k}^l}{\Delta t}&=\frac{1}{2}\Bigg[\frac{-b_{i,j,k+2}N_{i,j,k+2}^{l+1}+4b_{i,j,k+1}N_{i,j,k+1}^{l+1}
-3b_{i,j,k}N_{i,j,k}^{l+1}}{2\Delta E}\\
&+\frac{-b_{i,j,k+2}N_{i,j,k+2}^{l}+4b_{i,j,k+1}N_{i,j,k+1}^{l}-3b_{i,j,k}N_{i,j,k}^{l}}{2\Delta E}\Bigg].
\end{split}
\end{equation}
where $i$ and $j$ are the indexes of the spatial step in $r$ direction and $z$ direction respectively, while $k$ is the index of the energy step. The above equation can be reduced into
\begin{equation}
\begin{split}
N_{i,j,k}^{l+1}&=\left[\frac{-b_{i,j,k+2}N_{i,j,k+2}^{l+1}+4b_{i,j,k+1}N_{i,j,k+1}^{l+1}-3b_{i,j,k}N_{i,j,k}^{l+1}-b_{i,j,k+2}N_{i,j,k+2}^{l}+4b_{i,j,k+1}N_{i,j,k+1}^{l}}{4\Delta E}\Delta t+ N_{i,j,k}^l\right]\\
&\Big/\left[1+\frac{3b_{i,j,k}\Delta t}{4\Delta E} \right].
\end{split}
\end{equation}
Given the boundary condition $N_{i,j,k_{\rm max}}=0$ for any $l$, we can solve $N_{i,j,k}^{l+1}$ from $k=k_{\rm max}-1$ to $k=0$. Note that when adopting the Strang splitting scheme, one should replace $\Delta t$ by $\Delta t/2$.

To solve the diffusion equation, we first discretize the equation implementing the Crank-Nicolson scheme, and  obtain
\begin{equation}
\begin{split}
&\frac{N_{i,j,k}^{l+1}-N_{i,j,k}^l}{\Delta t}=\frac{D_{rr,k}}{2r_i}\left(\frac{N_{i+1,j,k}^l-N_{i-1,j,k}^l}{2\Delta r}+\frac{N_{i+1,j,k}^{l+1}-N_{i-1,j,k}^{l+1}}{2\Delta r}\right)\\
&+\frac{D_{rr,k}}{2}\left(\frac{N_{i+1,j,k}^l-2N_{i,j,k}^l+N_{i-1,j,k}^l}{\Delta r^2}+\frac{N_{i+1,j,k}^{l+1}-2N_{i,j,k}^{l+1}+N_{i-1,j,k}^{l+1}}{\Delta r^2} \right)\\
&+\frac{D_{zz,k}}{2}\left(\frac{N_{i,j+1,k}^l-2N_{i,j,k}^l+N_{i,j-1,k}^l}{\Delta z^2}+\frac{N_{i,j+1,k}^{l+1}-2N_{i,j,k}^{l+1}+N_{i,j-1,k}^{l+1}}{\Delta z^2} \right)
\end{split}
\end{equation}
By applying the ADI method, the above equation can be divided into
\begin{equation}
\begin{split}
\frac{N_{i,j,k}^{l+1/2}-N_{i,j,k}^l}{\Delta t}&=\frac{D_{rr,k}}{2r_i}\left(\frac{N_{i+1,j,k}^{l+1/2}-N_{i-1,j,k}^{l+1/2}}{2\Delta r}\right)
+\frac{D_{rr,k}}{2}\left(\frac{N_{i+1,j,k}^{l+1/2}-2N_{i,j,k}^{l+1/2}+N_{i-1,j,k}^{l+1/2}}{\Delta r^2} \right)\\
&+\frac{D_{zz,k}}{2}\left(\frac{N_{i,j+1,k}^l-2N_{i,j,k}^l+N_{i,j-1,k}^l}{\Delta z^2}\right)
\end{split}
\end{equation}
and
\begin{equation}
\begin{split}
\frac{N_{i,j,k}^{l+1}-N_{i,j,k}^{l+1/2}}{\Delta t}&=\frac{D_{rr,k}}{2r_i}\left(\frac{N_{i+1,j,k}^{l+1/2}-N_{i-1,j,k}^{l+1/2}}{2\Delta r}\right)
+\frac{D_{rr,k}}{2}\left(\frac{N_{i+1,j,k}^{l+1/2}-2N_{i,j,k}^{l+1/2}+N_{i-1,j,k}^{l+1/2}}{\Delta r^2} \right)\\
&+\frac{D_{zz,k}}{2}\left(\frac{N_{i,j+1,k}^{l+1}-2N_{i,j,k}^{l+1}+N_{i,j-1,k}^{l+1}}{\Delta z^2}\right)
\end{split}
\end{equation}
For the inner boundary condition where $i=0$ and/or $j=0$, we utilize the spatial symmetry, i.e., $N_{-1,j,k}=N_{1,j,k}$ and $N_{i,-1,k}=N_{i,1,k}$. We impose the CR density to be 0 at the outer boundary which is set to be $r=150\,$pc and/or $z=2.5\,$kpc.

\section*{Simulation of the turbulent magnetic field topology in ISM}
\begin{wrapfigure}{r}{0.5\textwidth}
\includegraphics[width=0.48\textwidth]{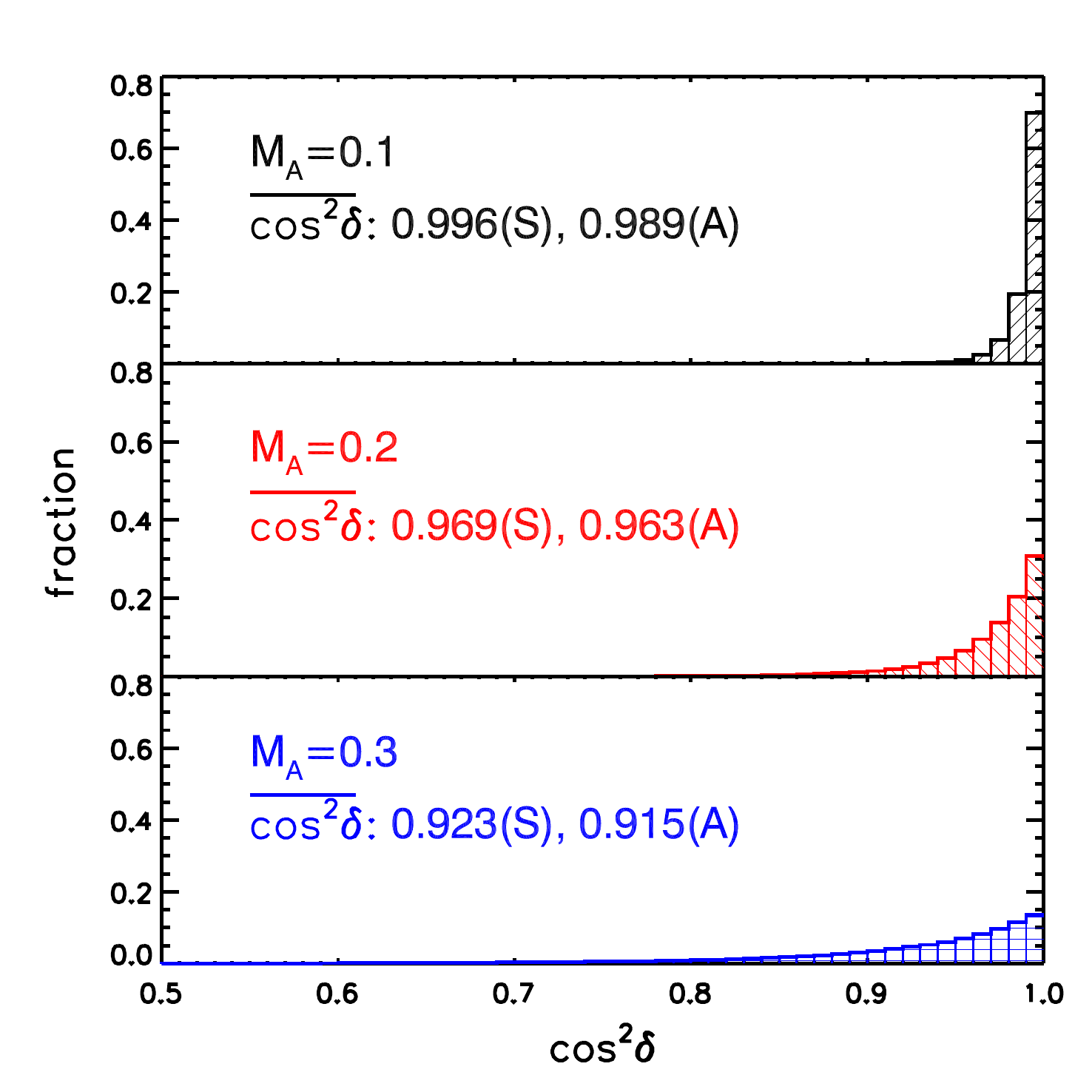}
\caption{Distribution of $\cos^2\delta$ for $M_A=0.1$ (top), $M_A=0.2$ (middle) and $M_A=0.3$ (bottom). The average value of $\cos^2\delta$ from simulation ('S') and from analytical estimation ('A') are also labelled.}
\end{wrapfigure}
The synchrotron radiation power highly dependent on the pitch angle of electrons that moving towards the observer with respect to the magnetic field direction.  This pitch angle, as derived in the main text, is mainly determined by the observer's viewing angle, i.e., the angle between the observer's line of sight and mean magnetic field direction of the interstellar medium (ISM) around the TeV halo, but the turbulence in the magnetic field would cause a small deviation of the local magnetic field direction from the mean magnetic field direction. This would increase the resulting synchrotron flux especially in the case of very small viewing angle. The angle of the deviation (denoted by $\delta$) were obtained from MHD turbulence simulations with different Alfv{\'e}nic Mach number $M_A$ [21-25].  Two state-of-the-art MHD codes (Pluto, {\url{http://plutocode.ph.unito.it}} and Pencil, {\url{http://pencil-code.nordita.org}}) for cross check purpose. The generated magnetic field topologies by these two codes are statistically identical in terms of the distribution of $\cos^2\delta$, which is the relevant quantity in calculating the synchrotron radiation. We show the distribution of $\cos^2\delta$ for $M_A=0.1, 0.2, 0.3$ in Fig.4. 
Another cross check for the generated $\theta$ comes from an analytic estimation of the average $\cos^2\theta$ [26], which reads $\overline{\cos^2\delta} \simeq 1/(1+M_A^2)$. The average value of $\cos^2\delta$ from simulation and analytical estimation are also labelled in Fig.~1. We can see that the differences between the simulation and the analytical estimations for all three $M_A$ are less than $1\%$.

\end{widetext}

\end{document}